\documentclass[conference]{IEEEtran}
\IEEEoverridecommandlockouts
\usepackage{cite}
\usepackage{amsmath,amssymb,amsfonts}
\usepackage{algorithmic}
\usepackage{graphicx}
\usepackage{textcomp}
\usepackage{xcolor}
\usepackage{array}
\usepackage{url}
\usepackage{lineno,hyperref}
\usepackage{color,framed}
\usepackage{color,xcolor}
\usepackage{algorithm}  
\usepackage{algorithmic}
\usepackage{threeparttable}
\usepackage{multirow}
\usepackage{longtable}
\usepackage{ulem}
\usepackage{array}
\usepackage{booktabs}
\usepackage{amsmath}
\usepackage{amssymb}
\usepackage{lineno,hyperref}
\usepackage{listings,xcolor}
\usepackage{courier}
\usepackage{pdflscape}
\usepackage{bbding}
\usepackage{adjustbox}
\usepackage{enumerate}

\usepackage{booktabs}
\newcommand{\tabincell}[2]{\begin{tabular}{@{}#1@{}}#2\end{tabular}}
\newcommand{\reffig}[1]{Fig.\ref{#1}}

\def\BibTeX{{\rm B\kern-.05em{\sc i\kern-.025em b}\kern-.08em
    T\kern-.1667em\lower.7ex\hbox{E}\kern-.125emX}}
\begin{document}

\title{NFTCert: NFT-Based Certificates With Online Payment Gateway}

\author{\IEEEauthorblockN{Xiongfei Zhao}
\IEEEauthorblockA{\textit{Department of Computer and Information Science} \\
\textit{University of Macau}\\
Macau \\
yb97480@umac.mo}
\and
\IEEEauthorblockN{Yain-Whar Si}
\IEEEauthorblockA{\textit{Department of Computer and Information Science} \\
\textit{University of Macau}\\
Macau \\
fstasp@um.edu.mo}
}

\maketitle

\begin{abstract}

Nowadays, academic certificates are still widely issued in paper format. Traditional certificate verification is a lengthy, manually intensive, and sometimes expensive process. In this paper, we propose a novel NFT-based certificate framework called NFTCert, which enables the establishment of links between a legitimate certificate and its owner through a Blockchain. In this paper, we describe the implementation of the NFTCert framework, including schema definition, minting, verification, and revocation of NFT-based certificates. We also introduce a payment gateway into the minting process, which enables NFTCert to be used by a wider audience. Therefore, participants of NFTCerts do not need to rely on cryptocurrency for transactions. All in all, the proposed framework is designed to achieve usability, authenticity, confidentiality, transparency, and availability properties when it is compared to existing Blockchain-based systems.

\end{abstract}

\begin{IEEEkeywords}
Blockchain, Smart Contract, NFT, Certificate
\end{IEEEkeywords}

\section{Introduction}

Non-fungible Token (NFT) is a cryptocurrency derivative of Ethereum smart contract and represents a unique digital identifier of ownership managed by Blockchain. NFT generally can be used to prove the authenticity and legitimacy of digital assets existing in or originating from the digital world, such as digital arts, ownership of physical assets, domain names, collections, event tickets, etc. Among them, the digital certificate in the form of NFT has brought many conveniences in issuance and legitimacy authentication. Many studies have introduced Blockchain-related technologies into certificate management scenarios, but the research on introducing NFT into the field of educational certificate management is still in a very early stage.

\begin{figure}[htbp]
        \makeatletter
        \def\@captype{figure}
        \makeatother
        \centering
        \includegraphics[width=0.38 \textwidth]{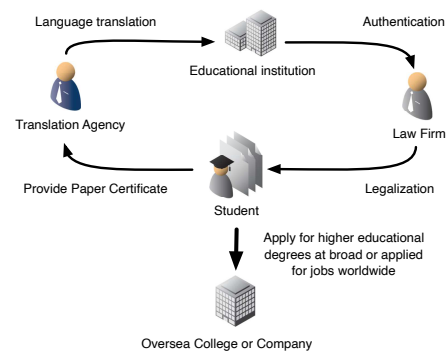}
        \setlength{\abovecaptionskip}{-0.1cm} 
        \caption{Certificate or certification authentication process without NFT}
        \label{CertificateWithoutNFT}
    \end{figure}

Nowadays, more and more international students have studied for higher educational degrees at broad or applied for jobs worldwide. According to the national bureau of statistics of China, in the year 2020, higher education graduates in China exceed 12.4 million \cite{graduatesnum}, they all receive a paper certificate for now. The number of higher education graduates in OECD and G20 countries is expected to surpass 300 million by 2030 \cite{be5514d7-en}. Among these students, those who apply for overseas study or a job are likely to be required to do language translations and international authentications or legalization regarding their original documents as a way to prove their authenticity (\reffig{CertificateWithoutNFT}). Besides, if students have multiple certificates, they need to repeat the above certification process for each paper certificate they hold and then send the certification information to overseas universities or companies. This cumbersome process leads to a lengthy process of ensuring the authenticity and legitimacy of certificates and involves considerable manual processing. Hence, the traditional education certificate system faces challenges from three main areas: the authenticity of educational institutions, certificate issuance, and certificate verification. To alleviate these problems, in this paper, we propose a novel NFT-based education certificate management framework called NFTCert.

    \vspace{-0.3cm}

NFTCert first adopts the private Blockchain architecture, and only real and verified participants can join. Therefore, educational institutions, overseas universities, or companies must obtain permission to join the private Blockchain to obtain certification-related services. Secondly, in terms of certificate issuance, NFTCert defines its own digital certificate data format, adds the signature of students' personal information and certificate information, and then issues the certificate to students and stores it in the digital wallet. Because a NFT certificate will not contain any students' personal information, it can protect the personal privacy of NFT certificate holders. Finally, during the certificate verification process, overseas universities or companies can clearly see which academic credentials students have based on the NFT stored in their digital wallets. Overseas universities or companies can first combine the personal information provided by students through their resumes, such as name, date of birth, place of birth, etc., with certificate information in the NFT to generate hash signatures. The ownership and authenticity of the certificate can be verified by comparing generated hash signature with the hash signature saved in the NFT.

Although managing certificates through NFT can bring many benefits, due to the current practice in NFT marketplace, participants must use cryptocurrency as a mean of payment to buy NFTs. Moreover, the high volatility of cryptocurrency often leads speculative cryptocurrency transactions, thus posing a threat to the current monetary system. Furthermore, cryptocurrency is often associated with illegal activities such as drug smuggling and money laundering. Therefore, many countries have completely banned financial institutions and payment companies from providing services related to cryptocurrency transactions, which makes it more challenging to adopt NFT in certificate management scenarios. 

In this paper, to reduce the reliance on cryptocurrency for transactions related to certificates, we integrate the online gateway into the NFT certificate issuance framework. By introducing a widely accepted and used online payment gateway as a payment method, our proposal can reduce the complexity for applicants to use NFTCert. Besides, our solution can be more widely used in regions where cryptocurrency is strictly regulated. We also compare NFTCert with existing Blockchain-based certificate management solutions, such as Blockcerts \cite{blockcerts}, EduCTX \cite{eductx}, BCDiploma \cite{BCDiploma}, Blockchain for Education \cite{grather2018Blockchain} and University of ZuricH Blockchain (UZHBC) \cite{UZHBC}. To the best of authors' knowledge, NFTCert proposed in this paper is the first NFT based framework for certificates management. Representing the educational certificate through NFTCert brings us many advantages. 
\begin{itemize}
    \item \textbf{\textsl{Usability}} Through the online payment gateway to pay for the cost of performing NFT-based certificate-related transactions, we can reduce the complexity of participation and expand the application region.
    \item \textbf{\textsl{Authentication}} NFT's token metadata and ownership can be publicly verified. It is easy for students to share their NFT wallets with overseas universities or companies to verify their academic achievements.
    \item \textbf{\textsl{Confidentiality}} NFTCert can ensure that the certificate information can be verified to belong to the student, on the premise that the student's personal information is not published to the Blockchain. 
    \item \textbf{\textsl{Transparency}} The activities of NFTs, such as minting, transferring are publicly accessible. It is easy for overseas universities or companies to identify which educational institution issue this NFT certificate to the applicant.
    \item \textbf{\textsl{Availability}} In the proposed framework, all certification information is stored in NFT. Students only need to record their NFT wallet address instead of keeping very digital certificates by themselves. The availability of Blockchain is typically high, therefore all the issued NFTs are always available for verification.
\end{itemize}

This paper is organized as follows. Section 2 reviews existing Blockchain-based certificate solutions. Section 3 gives the introduction of related concepts. Section 4 describes the proposed framework. In section 5, we evaluate the advantages of the proposed framework over existing solutions. The paper is concluded in Section 6.

\section{Related work}

Several institutes already developed Blockchain-based certificate systems to issue certificates. Typical Blockchain-based certificate system examples include Blockcerts \cite{blockcerts}, EduCTX \cite{eductx}, BCDiploma \cite{BCDiploma}, Blockchain for Education \cite{grather2018Blockchain} and University of ZuricH Blockchain (UZHBC) \cite{UZHBC}. Blockcerts is an initiative by the Massachusetts Institute of Technology (MIT). Blockcerts is an open standard for issuing and verifying Blockchain-based certificates. Digital certificates are registered on a Blockchain, and these certificates are accessible via an App termed Blockerts wallet. Students can share their verifiable, tamper-proof digital diplomas with overseas universities or companies.

EduCTX is another mature global higher education credit platform and ecosystem-based on Blockchain. Based on the open-source Ark Blockchain Platform, EduCTX tries to build a globally distributed peer-to-peer network. Higher Education Institute (EI) can join EduCTX Blockchain network as nodes with support and recognition with other HEI institutions. EduCTX is also based on the Concept of the European Credit Transfer and Accumulation System (ECTS), which will issue and manage ECTX tokens that represent the credits students have earned by completing courses such as ECTS. Students send their Blockchain wallet addresses and exchange scripts to overseas universities or companies for verification to present completed courses.

BCDiploma, Blockchain for Education, and UZHBC are all based on Ethereum to issue and verify diplomas. BCDiploma generates a unique Internet link that allows students to promote the authenticity of their diploma to the people with whom they share this link. Blockchain for Education introduces two smart contracts (IdentityMgmt and CertMgmt) to manage the public keys for certificate authorities and hashes of the certificates. Blockchain for Education also introduces IPFS to store certificate authorities profile and Basic Support for Cooperative Work (BSCW) document management system to store certificates and registers hash of the certificate in Blockchain. UZHBC stores the hash of the PDF document, which corresponds to the paper-based certificate. Overseas universities or companies can verify the certificate by checking the authenticity of the hash of the PDF document.

While these certificate systems have played a role in reducing fraudulent activities by taking advantage of Blockchain characteristics, they still have several limitations. For example, Blockcerts stores digital certificate's hash by conducting the costly transaction in Bitcoin. EduCTX only supports Higher Education accomplishments and does not support non-academic certificates. BCDiploma, Blockchain for Education, and UZHBC all store hashes of digital certificates on Ethereum. For BCDiploma, users need to use BCDT currency to pay for transactions. For UZHBC, it can only issue certificates for the University of Zurich departments. In contrast to these approaches, in this paper, we propose a novel NFT-based certificate management system, which not only overcomes the weaknesses of these systems but also introduces new features that are not supported by existing systems. Detailed comparison with existing systems is given in Table~\ref{CertificateManagementSystems}.

\section{Background}

In this section, we briefly introduce some of the terminologies that are going to be used in the proposed framework:

\textbf{\textsl{Private Blockchain}} Only authentic and verified participants can join a private Blockchain, usually implemented in an information-sensitive private business.

\textbf{\textsl{Ethereum Blockchain}} Ethereum \cite{ethereum.org} goes beyond just cryptocurrency transfer, and it also enables developers to deploy Turing-complete smart contract scripts.

\textbf{\textsl{Smart Contract}} Ethereum implements a complete Turing system to automatically move digital assets according to predefined arbitrary rules called smart contracts. The contents and conditions of execution are predetermined in smart contracts and will be automatically executed when the conditions are met \cite{ethereum.org}.

\textbf{\textsl{Digital Wallet}} Digital wallet is similar to a bank account, and it is a unique identifier for a user to send and receive the assets. Digital wallet is a fixed number of characters address which generated from a pair of public key and private key.

\textbf{\textsl{ERC20}} The ERC20 (Ethereum Request for Comments 20) \cite{EIP20}, introduces a standard for Fungible Tokens. ERC20 tokens are blockchain-based assets that each token is exactly the same (in type and value) as another token.
    
\textbf{\textsl{ERC721}} The ERC721 (Ethereum Request for Comments 721) \cite{EIP721}, introduces a standard for NFT. ERC721 token is unique and even 2 ERC721 token issued from same Smart Contract, they can have a different value. ERC721 helps us find a way to present distinctive details about an asset in the form of a token. 

\textbf{\textsl{Blockchain Oracle}} Blockchains and smart contracts cannot access off-chain data (data that is outside of the network). However, for many contractual agreements, it is vital to have relevant information from the outside world to execute the agreement.

\textbf{\textsl{Online Payment Gateway}} Online payment gateway such as Paypal \cite{PayPal}, Alipay \cite{Alipay}, Wechat Pay \cite{WeChatPay} are payment processing systems that authenticate the credentials, process securely recording and transmitting necessary details of the payment. A payment gateway issues a confirmation to all the involved stakeholders, enabling the transfer of money from the consumer's account to the merchant's account.

\section{NFTCert}

In this paper, we propose a novel NFT-based certificates management framework called NFTCert to prevent fraudulent entities from producing fake or illegitimate certificates. Overseas universities or companies can access student's educational certificates with the functionalities of NFT. In NFTCert, we propose an approach for using the traditional online payment to pay the fees incurred in NFTCert. By integrating NFT-based certificates with traditional online payment, we can reduce the complexity of using NFT-based certificates and allow stakeholders involved in this process to monetize the system. Our proposal is based on an isolated private Blockchain. Only authentic and verified educational institutions, overseas universities, and companies are able to join.

\begin{figure}[htbp]
    \makeatletter
    \def\@captype{figure}
    \makeatother
    \centering
    \includegraphics[width=0.45 \textwidth]{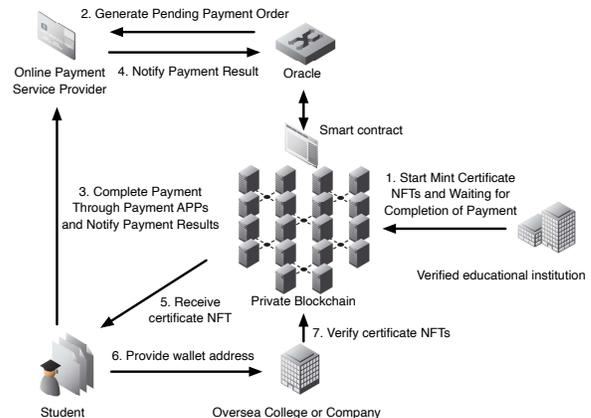}
        \setlength{\abovecaptionskip}{-0.2cm} 
    \caption{NFTCert certificate issuance and authentication diagram}
    \label{CertificateGeneral}
\end{figure}

As shown in \reffig{CertificateGeneral}, verified educational institutions join NFTCert's private Blockchain to mint NFT-based certificates for students. Each educational institute participates in this private Blockchain as nodes to maintain the integrity and creates trust among users. Certification fees are assessed for direct costs incurred in issuing a certificate, and educational institutions are charging these fees through an online payment gateway based on the following steps:
\begin{enumerate}[Step 1.]
    \item Verificated educational institution first mints certificate NFTs and waits for students to complete the payment before sending this NFT to student's digital wallet.
    \item Oracle will initiate a request to the online payment service provider to generate a pending payment order based on the smart contract request.
    \item After receiving the pending payment order, students will finish the payment through mainstream payment apps such as Paypal, Alipay, WeChat Pay, etc.
    \item The payment result will be forwarded to the smart contract through Oracle.
    \item After receiving the successful payment result, the smart contract will send the minted certificate NFT to the student's digital wallet.
    \item Student in possession of his/her NFT-based certificates can share the digital wallet address voluntarily with their counterparts. Colleges, employers, or any entity who needs their certificate info can easily verify the authenticity of the NFT-based certificates.
    \item Oversea universities or companies who are connected to the private Blockchain can access the stored NFT-based certificates. 
\end{enumerate}

\vspace{-0.4cm}

Next, we introduce four essential components of our proposed NFTCert system. The first component is the schema, which is based on the ECR-721 specification and describes the data standards or fields required for the certificate. The second component is the minting process, which creates a verifiable hash signature and issues the NFT-based certificate on Blockchain by NFTCert, which supports online payment. The third component is the verification process, which demonstrates how to base on student's info and NFT certificate info to verify the authenticity and ownership of the certificate. The last component is the revocation process, which describes actions to be made to revoke an NFTCert certificate. We do not consider trading of NFT based certificates in the proposed framework because these certificates are not supposed to be traded like other NFTs.

\subsection{Schema}

According to the ERC721 standard, we define a metadata JSON schema (\reffig{NFTCertMetadata}) to represent the certificate. The metadata JSON includes educational institution name, degree title, degree conferral date, certificate URI, and a hash signature:

\vspace{-0.4cm}
    \begin{figure}[htbp]
        \makeatletter
        \def\@captype{figure}
        \makeatother
        \flushleft
        \includegraphics[width=0.35\textwidth]{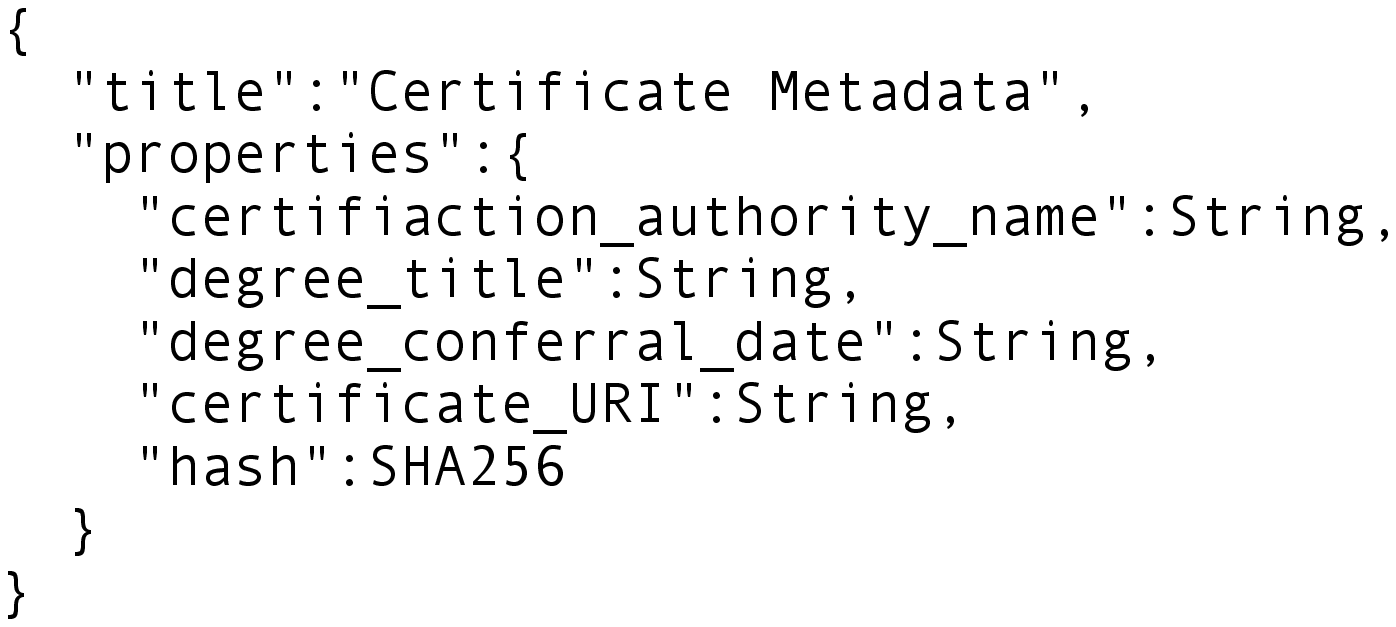}
        \setlength{\abovecaptionskip}{-0.3cm} 
        \caption{NFTCert metadata JSON schema}
        \label{NFTCertMetadata}
    \end{figure}
\vspace{-0.2cm}

The hash in \reffig{NFTCertMetadata} is populated as a SHA256 of student's name, date of birth, birthplace, nationality, educational institution name, degree title, degree conferral date, and certificate URI. Overseas universities or companies can use the NFT information with the student's names, date of birth, birthplace, nationality to verify whether the NFT-based certificate belongs to the applicant. Meanwhile, such mechanism can also prevent unauthorized disclosure of student's personal information.

The certificate\_URI in \reffig{NFTCertMetadata} is populated as a public URL pointing to the certificate image stored in the college's centralized web server. Alternatively, certificate\_URI can also be populated as an Interplanetary File System (IPFS) content-addressed hyperlink \cite{benet2014ipfs} that uses a hash value to uniquely describe the content itself. Because IPFS is a peer-to-peer distributed file system, certificate images can be cashed on different nodes across the network with no single point of failure.

\subsection{Minting process}

\begin{figure}[htbp]
    \makeatletter
    \def\@captype{figure}
    \makeatother
    \flushleft
    \includegraphics[width=0.44\textwidth]{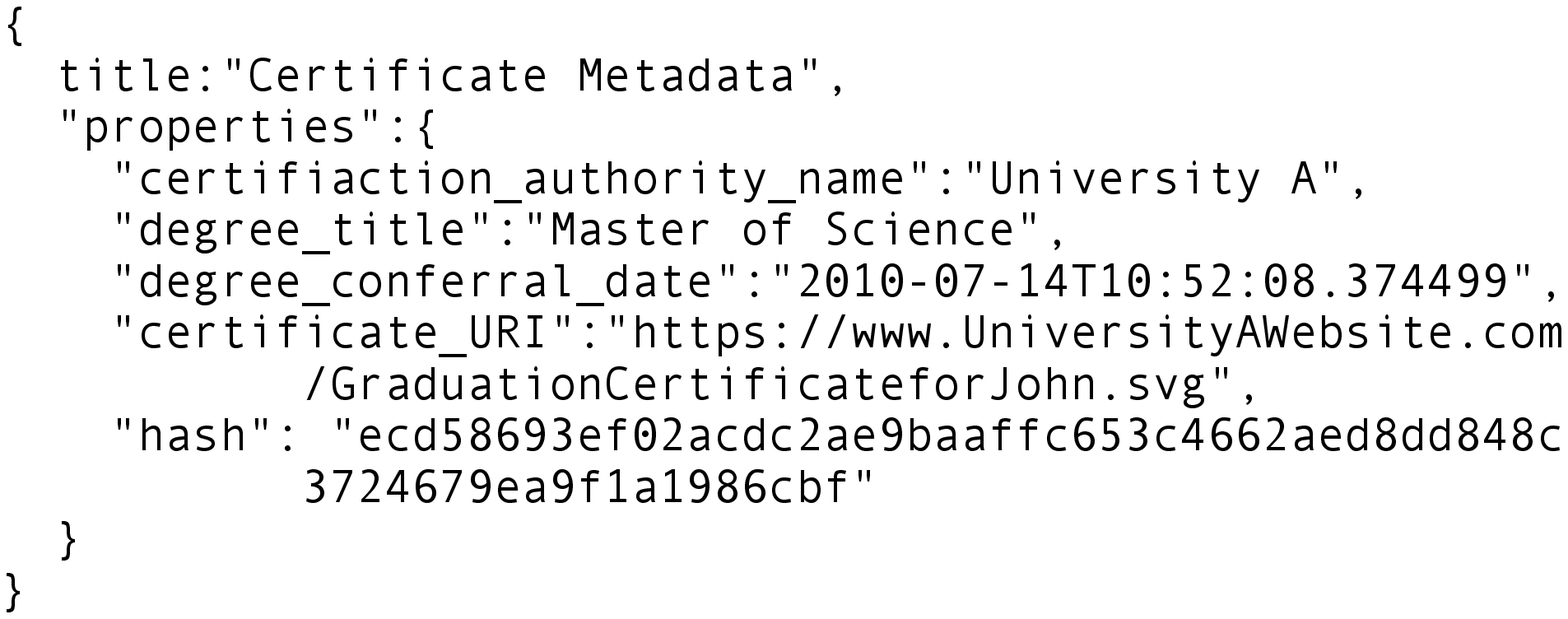}
    \setlength{\abovecaptionskip}{-0.2cm} 
    \caption{University A's NFTCert metadata JSON schema}
    \label{UniversityACertShm}
\end{figure}

To illustrate the mining process, we present University A's schematic graduation certificate for John in \reffig{UniversityACertPic}. After the certificate image is stored on the Web server or IPFS, the certificate URI in the JSON schema points to the certificate image associated with the NFT token. University A creates a metadata JSON schema (\reffig{UniversityACertShm}) to represent certificates based on student and educational institution information. This metadata JSON schema acts as a digital certificate for testifying that the recipient has successfully completed a particular course of study.

\vspace{-0.6cm}

Educational institution sends the JSON schema to Blockchain to generate NFT Token. Before NFT Token is transferred to the student's wallet, the recipient needs to pay the corresponding fee to the educational institution, which is usually a college or university. As previously explained in \reffig {CertificateGeneral}, the smart contract is responsible for interacting with traditional payment gateways through Oracle. The pricing information will be forwarded to the online payment service provider via Oracle, waiting for student to complete the payment.

\begin{figure}[H]
    \makeatletter
    \def\@captype{figure}
    \makeatother
    \flushleft
    \includegraphics[width=0.48\textwidth]{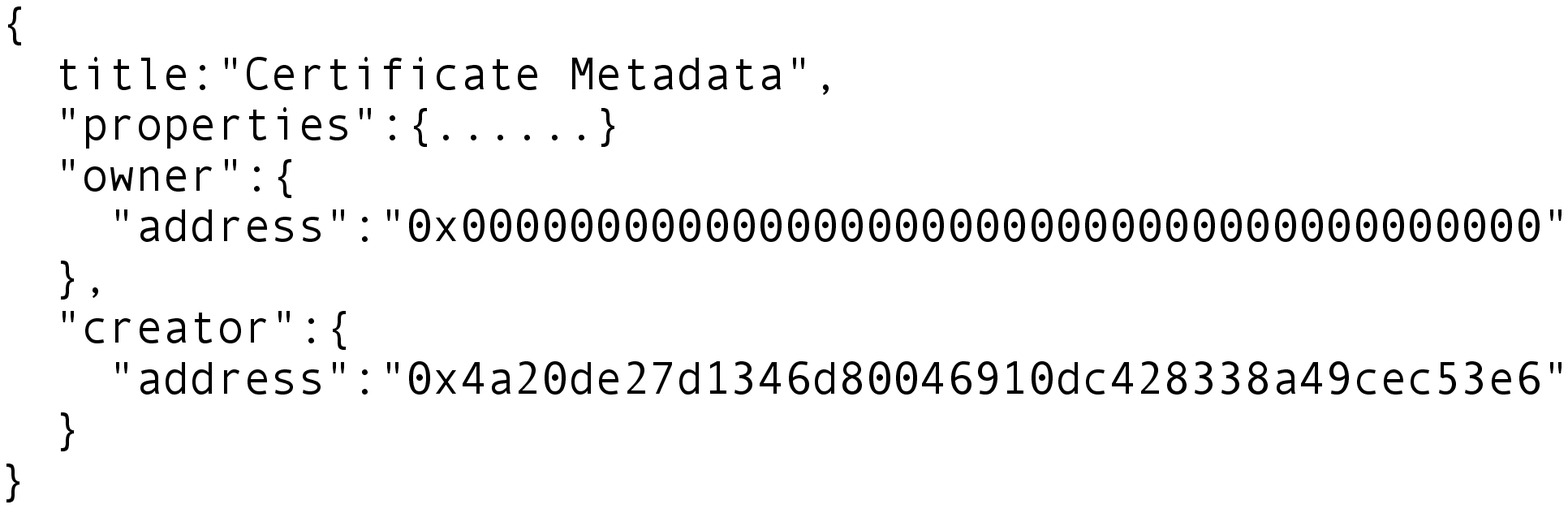}
    \setlength{\abovecaptionskip}{-0.6cm} 
    \caption{University A's NFTCert metadata JSON schema before payment}
    \label{UniversityACertBP}
\end{figure}

\reffig{UniversityACertBP} depicts University A's NFT-based certificate before the payment. Because the student has not finished the payment, this newly minted NFT-based certificate token's owner address is "0x00...000". The creator's address is University A's digital wallet address. University A then publishes the digital wallet address on its official website or government websites for public access. Overseas universities or companies can compare creator address in \reffig{UniversityACertBP} with the digital wallet address on University A's official website to identify the origin of this certificate.

\vspace{-0.4cm}
\vspace{-0.2cm}

\begin{figure}[H]
    \makeatletter
    \def\@captype{figure}
    \makeatother
    \flushleft
    \includegraphics[width=0.48\textwidth]{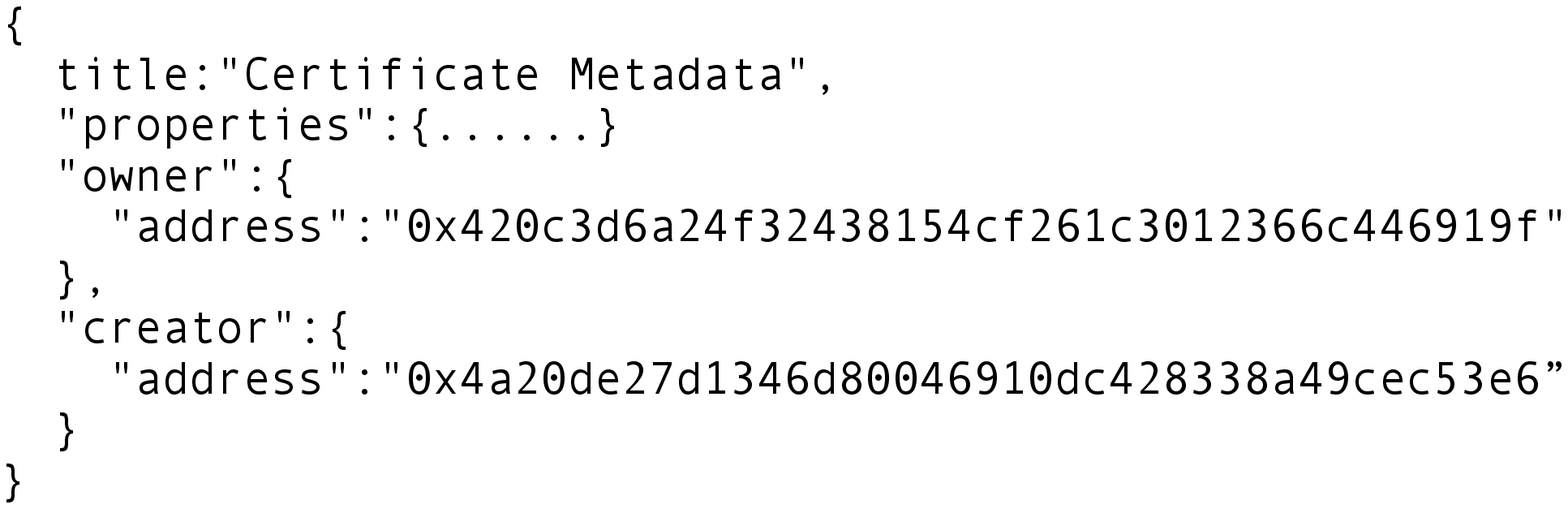}
    \setlength{\abovecaptionskip}{-0.6cm} 
    \caption{University A's NFTCert metadata JSON schema after payment}
    \label{UniversityACertAP}
\end{figure}

\reffig{UniversityACertAP} depicts the corresponding NFT-based certificate information in NFTCert after payment. The owner's address has been updated to the student's digital wallet address, which indicates that this certificate NFT has been transferred to the student's wallet. This finally ensures that the student received this certified certificate, and this certificate is ready to be verified by others in the future.

\vspace{-0.4cm}
\vspace{-0.1cm}

\subsection{Verification process}

In the past, paper certificates are used for authentications by prospective employers or educational institutes. These certificates can be easily lost or damaged. Moreover, paper certificates are often vulnerable to manipulation and forgery, and it is possible to make a counterfeit copy of the certificate outside of the institute. In some cases, a student just adds a fake certificate to his/her resume without achieving the real certificate. The cost of counterfeiting is low, and the cost of verification is high. In NFTCert, we propose a simple and inexpensive approach for the verification of the authenticity of the certificate. 

Since the NFT allows verification of its past history and owner, it is easy to verify that the previous owner is indeed the wallet address published by the educational institution on an official or government website. To determine whether the certificate belongs to the student, overseas universities or companies need to first extract the student's names, date of birth, birthplace, nationality from student's resume and then calculate the hash signature along with other certification-related information in the NFT metadata. By comparing signature calculation result with the hash signature in the NFT-based certificate, they can verify the authenticity of this certificate. Moreover, overseas universities or companies should be able to use the NFT schema certificate URI information to access the digital certificate image stored in an educational institution's server to further confirm the authenticity of the certificate. The steps for the verification process are as follow:

\begin{enumerate}[Step 1.]
  \item In \reffig{VerStep1}, overseas universities or companies first extract the following personal information from John's resume.

    \vspace{-0.4cm}
    \begin{figure}[H]
        \makeatletter
        \def\@captype{figure}
        \makeatother
        \flushleft
        \includegraphics[width=0.22\textwidth]{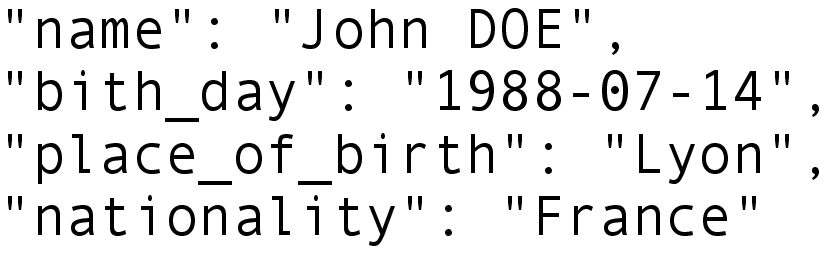}
        \setlength{\abovecaptionskip}{-0.2cm} 
        \caption{John's personal information extract form resume}
        \label{VerStep1}
    \end{figure}
    \vspace{-0.2cm}

    \item Overseas universities extracted other certification-related information (\reffig{VerStep2}) from University A's certificate metadata (\reffig{UniversityACertShm}).

    \vspace{-0.4cm}
    \begin{figure}[H]
        \makeatletter
        \def\@captype{figure}
        \makeatother
        \flushleft
        \includegraphics[width=0.44\textwidth]{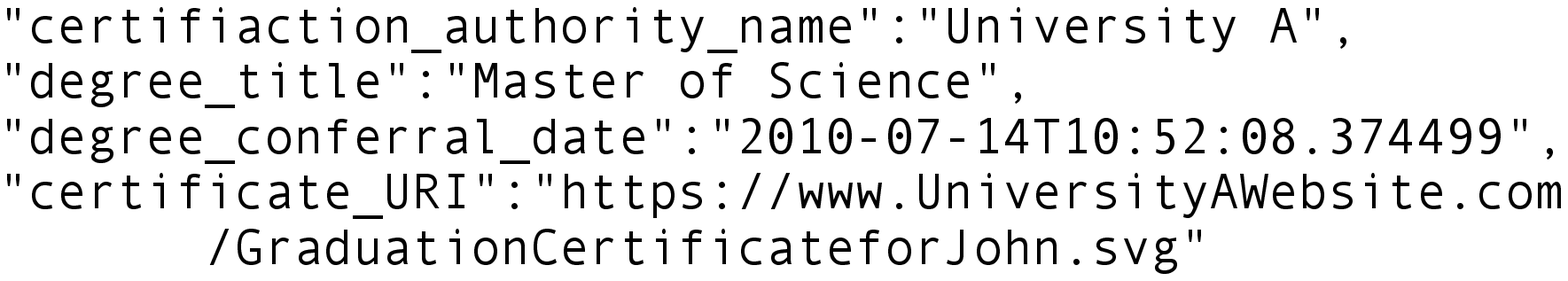}
        \setlength{\abovecaptionskip}{-0.2cm} 
        \caption{John's certificate information extract form John's wallet}
        \label{VerStep2}
    \end{figure}
    \vspace{-0.2cm}

    \item The hash signature (\reffig{VerStep3}) is calculated using the SHA256 cryptographic hash algorithm and then compared to the hash signature in the A University certificate metadata.

    \vspace{-0.4cm}
    \begin{figure}[H]
        \makeatletter
        \def\@captype{figure}
        \makeatother
        \flushleft
        \includegraphics[width=0.4\textwidth]{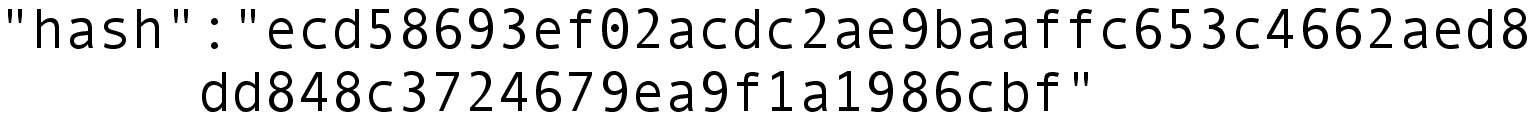}
        \setlength{\abovecaptionskip}{-0.2cm} 
        \caption{Hash signature calculated from John's personal and certificate information}
        \label{VerStep3}
    \end{figure}
    \vspace{-0.2cm}
\end{enumerate}
\vspace{-0.3cm}

\subsection{Revocation process}

For education entities to revoke certificates that were issued incorrectly, they could replace the certificate image file stored in the educational institution's server with a revoked declaration picture notice. Overseas universities or companies can confirm the validity of a certificate by assessing the NFT's certificate URI information.

\section{Discussion} 

Traditional education certificate systems face three major challenges: the authenticity of educational institutions, certificate issuance, and certificate verification. To ensure that NFTCert can alleviate the certificate management problem, we analyzed the advantages of NFTCert in solving each of the major challenges. 

In our proposal, we introduce private Blockchain. Only accredited educational institutions can join the private Blockchain and issue the certificate. Unaccredited certificate authorities or certificate mills can not issue the certificate. Meanwhile, activities of NFTs include minting and transferring, are publicly accessible. An educational institution can publish its NFT wallet address on authoritative channels such as official institution websites and government websites. The NFT wallet address of the educational institution on the website is interwoven with the NFT transfer record on the Blockchain, ensuring that each NFT was issued by the corresponding educational institution.


NFT-based certificates can have only one official owner. Since the Blockchain secures NFT-based certificates, no one can modify the record of ownership or duplicate a new NFT-based certificate into existence. With NFT enabled certificates, educational entities are able to mint certified NFT-based certificates as a digital token and transfer them to ones' digital wallet for future verification. Because NFT's token metadata and ownership are publicly accessible, we do not include any personal information in NFT's metadata. Instead, we add a hash field in NFT's metadata, and this hash is populated as a SHA256 of the student's names, date of birth, birthplace, nationality, and other certification-related information, include educational institution name, degree title, degree conferral date, and certificate URI.


Using NFT to record people's certificates in their own digital wallets could facilitate access to educational academic qualifications in a more efficient and transparent fashion. Overseas universities or enterprises could use people's NFT-based certificates to verify how many academic accomplishments this person has achieved. Because minting and transferring activities of NFTs are publicly accessible, it is easy to verify whether the corresponding educational institution issues certain NFT. When students study abroad or apply for a job, they will provide personal information in their resumes. Overseas universities or companies first extract the applicant's names, date of birth, birthplace, nationality from the resume and then obtain the hash signature along with other certification-related information in the NFT metadata. By comparing the SHA256 hash signature with the hash signature in the NFT certificate, they can determine whether the certificate belongs to the applicant.

To evaluate the effectiveness of NFTCert, we compare the proposed approach with existing state-of-the-art proposals. The comparison results are listed in Table~\ref{CertificateManagementSystems}. Although a number of Blockchain-based certificate management platforms have been developed to ease the burden of record checking and verification, our proposal still has superior advantages in many aspects. In particular, we propose to use traditional online payment to pay the fees incurred in the processing of NFT-based certificate, while allowing users to collect all their educational certificates in a digital wallet for others to browse.


    \vspace{-0.4cm}
\linespread{1}
\begin{table}[htbp]
\small
\caption{Comparison of the existing proposals in Blockchain based certificate management} 
\vspace{-0.3cm}
\centering 
\begin{threeparttable}
\begin{tabular}{p{2.9cm} c c c c c c c c} 
\hline\hline 
Proposal & 1 & 2 & 3 & 4 & 5 & 6 & 7 \\ [1.5ex] 
\hline  
\specialrule{0em}{6pt}{1pt}
Blockerts(2016) \cite{blockcerts} & \Checkmark & \XSolidBrush & \Checkmark & \Checkmark & \XSolidBrush & \XSolidBrush & \XSolidBrush \\ 
\specialrule{0em}{3pt}{1pt}
EduCTX(2018) \cite{eductx} & \XSolidBrush & \XSolidBrush & \XSolidBrush & \Checkmark  & \Checkmark & \XSolidBrush & \XSolidBrush \\
\specialrule{0em}{3pt}{1pt}
BCDiploma(2018) \cite{BCDiploma} & \Checkmark & \XSolidBrush & \XSolidBrush & \Checkmark & \XSolidBrush & \XSolidBrush & \XSolidBrush \\
\specialrule{0em}{3pt}{1pt}
\tabincell{l}{Blockchain for \\Education(2018) \cite{grather2018Blockchain}} & \Checkmark & \Checkmark & \XSolidBrush & \Checkmark & \XSolidBrush & \XSolidBrush & \XSolidBrush \\
\specialrule{0em}{3pt}{1pt}
UZHBC(2018) \cite{UZHBC} & \XSolidBrush & \XSolidBrush & \XSolidBrush & \Checkmark & \XSolidBrush & \XSolidBrush & \XSolidBrush \\
\specialrule{0em}{3pt}{1pt}
\tabincell{l}{NFTCert\\ (proposed approach)}& \Checkmark & \Checkmark & \Checkmark & \Checkmark & \Checkmark & \Checkmark & \Checkmark \\ [1ex] 
\hline 
\end{tabular}
\begin{tablenotes}
\footnotesize
\item 1: Support any type of certificate; 2: Accredited institutions; 3: Certificate revocation 4: Privacy of personal information; 5: No cryptocurrency involved; 6: Verifiable certificate information is stored solely on the Blockchain; 7: Student's certificates are collected in single digital wallet; \Checkmark Considered;  \XSolidBrush Not considered;
\end{tablenotes}
\end{threeparttable}
\label{CertificateManagementSystems} 
\end{table}

\vspace{-0.4cm}
\section{Conclusion}

In this paper, we propose a novel NFT-based certificate framework called NFTCert, which can be used to replace traditional paper-based certificates without relying on crypto-currency while maintaining usability, authentication, confidentiality, transparency, and availability properties. 

NFTs are non-fungible and unique, and the value of an NFT-based certificate does not equal to the value of another NFT-based certificate. With the increasing fraud and misuse of educational certificates, it is crucial to design a trustworthy and decentralized certification system to simplify the verification of the authenticity of certificates. Our solution takes advantage of NFT's unique and easy-to-verify features to make students' digital certificates tamper-proof in nature. 

In order to enable NFTCert to be used by a wider audience, we propose the solution of using the traditional online payment to pay the fees incurred in the processing of NFT-based certificates. Meanwhile, our proposal allows users to keep all educational certificates in one place for others to browse, reducing time and efforts spent by the institute in the verification of their certificates. When overseas universities or companies verify someone's educational experience, these organizations just need to look into his/her digital wallet, and verifiable NFT-based certificates could directly point to certificated educational certificates located in the educational institution's server.

As for the future work, we are planning to evaluate the efficiency of NFT casting when a large number of certificates are to be issued during the peak season. We are also planning to evaluate the reliability and stability of the proposed framework. In addition, we are investigating the possibility of introducing a novel incentive mechanism to attract more stakeholders to become NFTCert's nodes and to make the NFTCert framework more robust.

\section*{Acknowledgment}
This research was funded by the University of Macau (File no. MYRG2019-00136-FST).
 
\bibliographystyle{unsrt}
\bibliography{mybibfile}
\end{document}